\documentclass[article]{jss}

\usepackage{thumbpdf,lmodern}
\usepackage{booktabs}
\usepackage{amssymb,amsmath}
\usepackage{algorithm}
\usepackage[noend]{algpseudocode}
\usepackage{enumitem}

\usepackage{framed}

\newcommand{\fct}[1]{\code{#1()}}

\title{\textbf{IsoCheck}: An \proglang{R} Package to check Isomorphism for Two-level Factorial Designs with Randomization Restrictions}
\Plaintitle{}
\Shorttitle{IsoCheck: \proglang{R} package for Isomorphism check}
\author{Pratishtha Batra \\ IIM Indore 
	\And Neil A. Spencer \\ Harvard University 
	\And Pritam Ranjan \\ IIM Indore}
\Plainauthor{Pratishtha Batra, Neil Spencer, Pritam Ranjan}

\Abstract{
Factorial designs are often used in various industrial and sociological experiments to identify significant factors and factor combinations that may affect the process response. In the statistics literature, several studies have investigated the analysis, construction, and isomorphism of factorial and fractional factorial designs. When there are multiple choices for a design, it is helpful to have an easy-to-use tool for identifying which are distinct, and which of those can be efficiently analyzed/has good theoretical properties. For this task, we present an \proglang{R} library called \pkg{IsoCheck} that checks the isomorphism of multi-stage $2^n$ factorial experiments with randomization restrictions. Through representing the factors and their combinations as a finite projective geometry,  \pkg{IsoCheck} recasts the problem of searching over all possible relabelings as a search over collineations, then exploits projective geometric properties of the space to make the search much more efficient. Furthermore, a bitstring representation of the factorial effects is used to characterize all possible rearrangements of designs, thus facilitating quick comparisons after relabeling. We present several examples with \proglang{R} code to illustrate the usage of the main functions in \pkg{IsoCheck}. Besides checking equivalence and isomorphism of $2^n$ multi-stage factorial designs, we demonstrate how the functions of the package can be used to create a catalog of all non-isomorphic designs, and subsequently rank these designs based on a suitably defined ranking criterion.  \pkg{IsoCheck} is free software and distributed under the General Public License and available from the Comprehensive \proglang{R} Archive Network.
}


\begin{document}


\section[Introduction]{Introduction} \label{sec:intro}

Factorial experiments are commonly used for significance assessment of factors in a wide spectrum of scientific and industrial applications including agriculture, engineering, and sociological phenomena. In many such experiments, complete randomization of the trials happens to be impractical for a variety of reasons. For instance, changing some factors can be prohibitively expensive, the trials may need to be partitioned into homogeneous batches at each stage, or the experimental units may need to be processed repeatedly in batches. To address these issues, randomization restrictions are imposed on the experimental trials. Classical designs for such factorial experiments include block designs, nested designs, split-plot designs, strip-plot designs, split-lot designs, and combinations thereof. A plethora of research articles have been published till date on the development of theoretical results and efficient algorithms to construct and analyze such designs.  Some landmark papers can be traced back to the exploration of completely randomized designs and randomized block designs by \cite{Fisher1942} and \cite{yates1937}.  Later, experimenters investigated several factorial designs with multi-stage randomization restrictions. For instance, \cite{addelman1964}, \cite{bingham_sitter_1999} and many others investigated the split-plot designs; \cite{miller_1997} developed the strip-plot designs for measuring wrinkles in a laundry experiment; blocked designs were investigated by \cite{bisgaard_1994}; \cite{mee_bates_1998} and Butler (2004) presented some fundamental results on split-lot designs for analyzing the fabrication of integrated circuits using silicon wafers. \\

One important aspect in this area of research focuses on finding \emph{distinct} designs that have useful theoretical properties and are easy to analyze. An essential task in such endeavors is to check for the isomorphism between different designs. In a completely randomized setup, two designs are said to be isomorphic if one can be obtained by applying relabeling and rearrangement of the factors and factor combinations on the other design. \cite{franklin_bailey_1977} developed theoretical results and algorithms for fractional factorial  completely randomized designs. For factorial designs with randomization restrictions, isomorphic designs allow restricted rearrangement of the factorial effects within the subspace that  characterizes the randomization defining factors.  \cite{bingham_sitter_1999} extended the results of \cite{franklin_bailey_1977}   for split-plot designs. In a multi-stage factorial design with randomization restrictions, there are additional restrictions on the within block and between block relabeling and rearrangement (see \cite{spencer_etal_2019}). Several researchers have also focused on proposing innovative design ranking criteria to sort through these non-isomorphic designs and create catalogs of good designs for ready usage by practitioners.  Popular design ranking criteria are maximum resolution, maximize the number of clear 2-factor interactions, minimum aberration, etc. \\

Over the years, a few methodologies have been developed that unify different factorial designs with randomization restrictions.  \cite{nelder_1965a, nelder_1965b} developed a unified theory to analyze such experiments through simple block structures. This idea was further generalized to orthogonal block structures through association schemes by \cite{speed_bailey_1982} and \cite{tjur_1984}. More details can be found in \cite{bailey2004association} and \cite{cheng2011paper}. \cite{bingham2008factorial} introduced the idea of randomization defining factors, which was further taken-up by \cite{ranjan2009existence} and \cite{ranjan2010stars} to develop Projective Geometry (PG) formulation based on the overlapping pattern of the randomization defining contrast subspaces (referred to as RDCSSs). \\

The key idea behind this RDCSS-based unified theory is to realise that the set of all factorial effects of a $q^n$ factorial design constitute an $(n-1)$-dimensional finite projective geometry $\mathcal{P}_{n} := PG(n-1, q)$ over $GF(q)$. The randomization restrictions can be characterized by a projective subspace of  $\mathcal{P}_{n}$. Such multi-stage factorial designs have been constructed using two geometric structures called ``spread" and ``star". The former partitions $\mathcal{P}_n$ into disjoint RDCSSs, whereas the latter is a set of RDCSSs that forms a cover of $\mathcal{P}_n$ with a common overlap. Ranjan et al. have also established that the construction and analysis of classical factorial designs with randomization restrictions like split-plot, strip-plot, split-lot, etc.,  can be derived from carefully chosen sets of RDCSSs of $\mathcal{P}_{n}$.  Recently, \cite{spencer_etal_2019} developed isomorphism check algorithms for $2^n$ multi-stage factorial designs based on spreads and stars. \\

This paper presents the \proglang{R} \citep{R-cran} package \pkg{IsoCheck} which assumes the RDCSS-based structure of the factorial designs and efficiently implement the algorithms of Spencer et al. to check the isomorphism. The important highlights of \pkg{IsoCheck} package are functions to (a) generate a relabelling and rearrangement of a given spread/star of $\mathcal{P}_n$ via a proper collineation matrix; (b) check equivalence of two spreads (stars); (c) check isomorphism between two spreads (stars); and (d) find a spread embedded in the star-based design. \\
     
Section~2 of the paper provides necessary theoretical background on factorial designs, foundations of projective geometry $\mathcal{P}_n=PG(n-1, 2)$, including balanced spreads and covering stars of $\mathcal{P}_n$, and algorithms for isomorphism check (developed by \cite{spencer_etal_2019}). Section~3 outlines the usage and arguments of important functions of \pkg{IsoCheck}. In Section~4, we use several examples to present detailed illustration of how isomorphism check can be performed for both spread and star based designs. Section~5 illustrates how the functions of  package can also be used to construct a catalog of all non-isomorphic designs, and subsequently rank these designs based on some suitably defined ranking criterion. Of course, it assumes that one can list out all possible designs with a given specification (i.e., number of factors, number of stages, and sizes of each RDCSS). Section~6 concludes the article with a few remarks.

\section{Methodology} \label{sec:models}

In a $2^n$ factorial experiment, suppose the $n$ basic factors (main effects), $F_{i}$, $i= 1, \dots, n$, are expressed as the $n$ canonical vectors, where $F_i$ has $n-1$ zeros and exactly one $1$ at the $i$-th location. These $n$ canonical vectors along with their interactions constitute an $n$-dimensional vector space $V$ over $GF(2)$. The set of all non-null elements of $V$ form a \emph{$(n-1)$-dimensional projective space over $GF(2)$}, denoted by $\mathcal{P}_{n} := PG(n-1, 2)$ \citep{bose1947mathematical}. For a given stage of a multi-stage / multi-stratum factorial design (e.g., split-plot, strip-plot, split-lot, etc.), the randomization restrictions can be characterized by a projective subspace of $\mathcal{P}_{n}$. In the context of projective geometry, such a subspace is called a flat, whereas in the design of experiments language, we refer to this as a \emph{randomization defining contrast subspace} (RDCSS), which is in the same spirit as \emph{fraction defining contrast subgroup} (FDCS) used to characterize traditional fractional factorial designs \citep{bingham_sitter_1999}. A $(t - 1)$- flat of $\mathcal{P}_{n}$ is a set of $2^{t} - 1$ non-null effects spanned by $t$ linearly independent effects of $\mathcal{P}_{n}$. \\

The multi-stage RDCSS-based designs proposed by \cite{ranjan2009existence} and \cite{ranjan2010stars} are primarily derived from two geometric structures called spreads and stars of $\mathcal{P}_n$, which are essentially the set of distinct flats of $\mathcal{P}_n$ that also form a cover of $\mathcal{P}_n$. In a nutshell, a spread is a collection of disjoint flats, whereas a star is a set of distinct flats that allow a common overlap. Ranjan et al. proposed several multi-stage factorial designs by exploiting the sizes of flats and their overlaps. Recently, \cite{spencer_etal_2019} has developed algorithms for checking isomorphism between a smaller class of balanced spread- and balanced star- based designs.  The \proglang{R} library (\pkg{IsoCheck}) presented here facilitates easy implementation of these isomorphism check algorithms.\\

\textbf{Definition :} For $1 \leq t \leq n$, a balanced $(t - 1)$-spread $\psi$ of $\mathcal{P}_{n} := PG(n-1, 2)$ is a set of $(t - 1)$-flats of $\mathcal{P}_{n}$ which partitions $\mathcal{P}_{n}$. \\

In other words, a \emph{spread} of $\mathcal{P}_{n}$ is a set of projective subspaces of $\mathcal{P}_{n}$ that covers $\mathcal{P}_{n}$ i.e., $\psi$ consists of all disjoint flats that includes every element of $\mathcal{P}_{n}$.  A $(t - 1)$-spread of $\mathcal{P}_{n}$ contains $\mu = (2^{n}-1)/(2^{t}-1) $ distinct $(t - 1)$-flats that can be used to construct the RDCSSs for different stages of randomization. A necessary and sufficient condition for the existence of a balanced $(t - 1)$-spread is that $t$ divides $n$ \citep{andre1954nicht}.\\

If $t$ does not divide $n$, then either a partial $( t - 1)$-spread or an overlapping set of RDCSSs have to be used for design construction. \cite{ranjan2009existence} presents some results on design construction using partial spreads. However, we restrict to the algorithms on the isomorphism of complete spread based designs.  In cases when overlap between two $(t-1)$-flats (i.e., RDCSSs) is unavoidable, we focus on balanced star-based designs in $\mathcal{P}_n$.\\

\textbf{Definition :} For $0 \leq t_{0} < t < n$, a \emph{balanced star} of $\mathcal{P}_{n}$ is a set of $\mu$ $(t- 1)$-flats and a $(t_{0} -1)$-flat in $\mathcal{P}_{n}$, such that the intersection of any two of the $\mu$ flats is the  $(t_{0} -1)$-flat. \\

Such a star, denoted by $\Omega = St(n, \mu, t, t_{0})$, consists of
$\mu$ distinct $(t- 1)$-flats called rays and a $(t_{0} -1)$-flat known as the nucleus. The covering star $ St(n, \mu, t, t_{0})$ provides $ \mu = (2^{n-t_0}-1)/(2^{t-t_0}-1)$ overlapping RDCSSs of size $2^t - 1$ each.\\

A balanced spread can be constructed using the cyclic approach of \cite{hirschfeld1998projective}. The non-zero elements of $GF(2^n)$ can be written as $\lbrace\omega^0, \omega^1, \dots, \omega^{2^n -2}\rbrace$, where $\omega$ is a primitive element, and $\omega^i = \alpha_{0} \omega^{0} + \alpha_{1} \omega^{1} + \dots + \alpha_{n-1} \omega^{n-1}$, for $0 \leq i \leq 2^{n}-2$, correspond to the vector representation $( \alpha_0, \dots , \alpha_{n-1})$ of elements in $\mathcal{P}_{n}$. Let $t$ be an integer where $t$ divides $n$. If we want to construct a $(t - 1)$-spread $\psi$ of $\mathcal{P}_{n}$, we write the non-zero elements of $GF(2^n)$ in cycles of length $\mu$, where $\mu = (2^n -1)/(2^t -1)$ . \cite{hirschfeld1998projective} showed that the $f_i's$ are $(t - 1)$-flats as shown in Table 1 and $\phi = \lbrace f_1, \dots , f_\mu\rbrace$ partitions the set of all non-zero elements of $GF(2^n)$, i.e., $\psi$ is a $(t - 1)$-spread of $\mathcal{P}_{n}$. \\

\begin{table}[!h]
\caption{The elements of $GF(2^n)$ in cycles of length $\mu$}
\label{Table 1}
\begin{center}
\begin{tabular}{@{}cccc@{}}
\toprule
$f_{1}$ & $f_{2}$ & $\dots$ & $f_{\mu}$  \\ \midrule
$\omega^0$ & $\omega^1$ & $\dots$ & $\omega^{\mu - 1}$ \\
$\omega^{\mu}$ & $\omega^{\mu + 1}$ & $\dots$ & $\omega^{2\mu - 1}$  \\
$\vdots$ & $\vdots$ & $\ddots$ & $\vdots$  \\
$\omega^{2^n - \mu -1}$& $\omega^{2^n - \mu}$ & $\dots$ & $\omega^{2^n - 2}$  \\ \midrule
\end{tabular}
\end{center}
\end{table}
 
 As a quick example, Table~2 shows a cyclic spread of $\mathcal{P}_4=PG(3,2)$ using the primitive polynomial $\omega^4 + \omega + 1$. One can use these five $1$-flats for constructing up to five RDCSSs of a multi-stage factorial design.\\

 \begin{table}[h!]\centering
 	\caption{A $1$-spread of $PG(3,2)$ obtained via the primitive polynomial $\omega^4 + \omega + 1$.}
\begin{tabular}{ccccc}
	\toprule
	$f_{1}$ & $f_{2}$ & $f_{3}$ & $f_{4}$ & $f_{5}$ 
	\\ \cmidrule(r){1-5} 
	D & C & B & A & CD \\
	BC & AB & ACD & BD & AC  \\
	BCD & ABC & ABCD & ABD & AD \\
	\bottomrule
\end{tabular}
\end{table}

We can also construct a balanced covering star $\Omega = St(n, \mu, t, t_{0})$ using such cyclic spreads as shown in Table~1. Let $ \lbrace f_1, \dots , f_\mu\rbrace$ denote the flats consistuting a $(t - t_0 -1)$-spread $\psi$ of $\mathcal{P}_{n - t_0}$. Then there exists a $(t_0 - 1)$-flat $\pi$ in $\mathcal{P}_{n} \setminus \mathcal{P}_{n - t_0}$ such that $f_{i}^{*} = \langle f_{i} , \pi\rangle$ and  $ \lbrace f_{1}^{*}, \dots , f_{\mu}^{*}\rbrace$ form a covering star $\Omega$ of $\mathcal{P}_{n}$. We denote such stars as $\Omega = \psi \times \pi$.\\

We now return to the necessary theoretical details for discussing isomorphism between two designs $d_1$ and $d_2$. Checking isomorphism between two multi-stage factorial designs requires essentially two major checks: (a) rearrangement of factorial effects within each RDCSS and the rearrangement / renaming of RDCSSs themselves, and (b) relabelling of factorial effects. The first step simply amounts to checking the equality of $d_1$ and $d_2$ as sets, and if satisfied, we refer to $d_1$ and $d_2$ as equivalent.\\

\textbf{Definition :} Let $d_1$ and $d_2$ be two $2^n$ multi-stage factorial designs with $\mu$ stages of randomization, and let the respective sets of RDCSSs be denoted by $d_1 = \lbrace f_{1} , f_{2} , \dots, f_{\mu}\rbrace$ and $d_2 = \lbrace g_{1} , g_{2} , \dots, g_{\mu}\rbrace$. Then $d_{1}$ is said to be \emph{equivalent} to $d_{2}$ (denoted by $d_1 \equiv d_2$) if and only if, for every $f_{i} \in d_{1}$, there is a unique $g_{j} \in d_{2}$ such that $\lbrace f_{i} \rbrace = \lbrace g_{j} \rbrace$ (set equality), for $1 \leq i, j \leq \mu$.\\

Let $\mathcal{E}(d_{1})$ denote the equivalence class of $d_{1}$, and if $\mid f_{i} \mid = 2^t - 1$ for $f_i \in d_1$, then, 
$$\mid \mathcal{E}(d_{1}) \mid = \mu ! [(2^t - 1)!]^{\mu}.$$
For instance, the size of equivalence class for the design presented in Table~2 is $5!((3!)^5)= 933120$. When $n$ is large, checking the equivalence of two designs by naively iterating through the entire equivalence class is too computationally intensive. However, the computational burden can be reduced through a combination of sorting and \emph{bitstring representation} scheme.\\

Every element of $\mathcal{P}_{n}$ can be represented as a unique binary string of $2^{n} - 1$ bits with exactly one non-zero entry. In this representation, the contents of any RDCSS $f$ can now be identified by the sum of the bitstring representations of its elements. For example, following the Yates Order \citep{box1978statistics} of $\mathcal{P}_{3}$, the bitstring representations of the elements of $\mathcal{P}_{3}= \lbrace A, B, AB, C, AC, BC, ABC\rbrace$ are $A \rightarrow 1000000$, $B \rightarrow 0100000$, $\dots$, $ABC \rightarrow 0000001$. Then, $f = \lbrace AB, AC, BC\rbrace$ can be uniquely represented by the bitstring representation $0010110$. As a result, the equivalence check between two designs $d_1$ and $d_2$ simplifies to only comparing two sets of $\mu$ bitstrings, which is much cheaper than having to compare $d_1$ to all possible rearrangements of $d_2$.\\

The second major step towards isomorphism check is to go through all possible relabelings of the factors and factor combinations in $d_1$ and then check for equivalence between the relabeled $d_1$ and $d_2$. In projective geometry literature, relabeling is typically done by applying a collineation. \\

\textbf{Definition :} A collineation of $\mathcal{P}_{n}=PG(n-1,2)$ is a mapping of the points from $\mathcal{P}_{n}$ to $\mathcal{P}_{n}$ such that $(t - 1)$-flats gets mapped to $(t - 1)$-flats for all $1 \leq t \leq n$.\\

A collineation of $\mathcal{P}_{n}$ can be characterized by an $n \times n$ matrix  $\mathcal{C}$ over $GF(2)$, referred to as the collineation matrix \citep{batten1997combinatorics}, where the $j$-th column of $\mathcal{C}$ is the image of the basic factor $F_j$ (i.e., the effect in $\mathcal{P}_{n}$ that $F_j$ gets mapped to). {For instance, a $3 \times 3$ collineation matrix for $PG(2,2)$ is 
$$
	\mathcal{C} =   \begin{pmatrix} 
	1& 0 & 0\\
	0& 1 & 1\\
	0& 0 & 1\\
	\end{pmatrix}
$$
	which relabels the basic factors as $A \rightarrow A$, $B \rightarrow B$ and $C \rightarrow BC$.\\

\textbf{Definition :} Two $2^n$ RDCSS-based designs $d_1 = \lbrace f_{1} , f_{2} , \dots, f_{\mu}\rbrace$ and $d_2 = \lbrace g_{1} , g_{2} , \dots, g_{\mu}\rbrace$ are said to be \emph{isomorphic} (denoted by $d_1 \cong d_2$) if and only if there exists a collineation $\mathcal{C}$ over $\mathcal{P}_{n}$ such that $\mathcal{C}(d_1) \equiv d_2$. In this case, we say that $\mathcal{C}$ is an isomorphism establishing collineation (IEC) from $d_1$ to $d_2$. \\

We now briefly describe the search algorithm for isomorphism check between two spread- (star-) based designs, proposed by \cite{spencer_etal_2019}.  \\

\subsection{Isomorphism Check for Spread-based Designs}
For checking the isomorphism between two $(t-1)$-spread based multi-stage factorial designs characterized by $d_1 = \lbrace f_{1} , f_{2} , \dots, f_{\mu}\rbrace$ and $d_2 = \lbrace g_{1} , g_{2} , \dots, g_{\mu}\rbrace$ in $\mathcal{P}_{n}$, we follow a systematic approach to search for the IEC by iterating through the candidate collineations. First, choose $l_0 = n/t$ out of $\mu$ RDCSSs in $d_1$ that can generate a basis $\lbrace x_1, \dots, x_n\rbrace$ for $\mathcal{P}_{n}$ such that each of the $l_0$ RDCSSs contributed $t$ independent points to the basis. Then, construct a collineation matrix $\mathcal{C}_{x,B}$ to transform the $x$’s to a canonical basis with basic factors,$\lbrace A,B, \dots \rbrace$. Note that the equivalence of $\mathcal{C}_{x,B}(d_1)$ and $d_2$ implies the isomorphism between $d_1$ and $d_2$. Now, iterate through all possible sets of $l_0$ RDCSSs from $d_2$ to construct the basis set$\lbrace y_1, \dots , y_n \rbrace$ in $d_2$ and the corresponding collineation matrix $\mathcal{C}_{B, y}$ that maps the canonical basis to $y$’s. If $\mathcal{C}_{B, y}(\mathcal{C}_{x,B}(d_1))$ is equivalent to $d_2$, then we have found the IEC. The step-by-step algorithm from \cite{spencer_etal_2019} is now presented for convenience of the readers.\\

\begin{algorithm}[h!]
\caption{Isomorphism check between two $(t-1)$-spread based designs in $\mathcal{P}_{n}$}
\begin{algorithmic}[1]

\item Choose $l_0 = n/t$ out of $\mu$ RDCSSs from  $d_1 = \lbrace f_{1} , f_{2} , \dots, f_{\mu}\rbrace$, such that $\langle\cup_{i=1}^{l_0} f_{u_i}\rangle = \mathcal{P}_{n}$. Let these be $\lbrace f_{u_1} , f_{u_2} , \dots, f_{u_{l_0}}\rbrace$ for $1 \leq u_1 , u_2 , \dots , u_{l_0} \leq \mu$   .

\item For $i= 1,\dots, l_0$,  specify $ \lbrace x_{i,1}, \dots, x_{i,t}\rbrace$ $\in f_{u_{i}}$ such that $f_{u_{i}}=  \langle x_{i,1}, \dots, x_{i,t}\rangle $.

\item Construct the collineation matrix $\mathcal{C}_{x,B}$ which maps the $x$'s in Step 2 to the canonical basis$\lbrace A, B, \dots \rbrace$ such that each $x_{i,j}$ is mapped to $F_{t(i-1))+j}$.

\item 
\begin{enumerate}[label=(\alph*)] 

\item Choose $l_0$ out of $\mu$ RDCSSs from  $d_2 = \lbrace g_{1} , g_{2} , \dots, g_{\mu} \rbrace$ for mapping $f_{u_i}$'s  (note that ordering is important). Let that be $\lbrace g_{v_1} , g_{v_2} , \dots, g_{v_{l_0}}\rbrace$ for $1 \leq v_1 , v_2 , \dots , v_{l_0} \leq \mu$  . If $\vert \langle g_{v_1} \cup \dots  \cup g_{v_{l_0}} \rangle \vert <  2^{n} - 1 $, proceed to the next choice for $\lbrace g_{v_1} , g_{v_2} , \dots, g_{v_{l_0}}\rbrace$. 

\item For $i= 1,\dots,l_0$, choose a  $ \lbrace y_{i,1}, \dots, y_{i,t}\rbrace$ $\in g_{v_{i}}$ such that $g_{v_{i}}=  \langle y_{i,1}, \dots, y_{i,t}\rangle $.

\item Choose one of the $(l_0)!$ permutations of the elements $1,\dots,l_0$, say $\sigma_k$, for $k = 1,\dots,(l_0)!$.

\item Construct $\mathcal{C}_{B,y}$ which maps the canonical basis elements to $y$'s such that $F_{t(i-1))+j}$ is mapped to $y_{\sigma_k(i), j}$.

\item If $\mathcal{C}_{B, y}(\mathcal{C}_{x,B}(d_1))$ is equivalent to $d_2$,then $d_1 \cong d_2$, and report $\mathcal{C} = \mathcal{C}_{B, y} . \mathcal{C}_{x,B}$ as an IEC and exit; otherwise, continue.

\item Go to Step 4(c) and choose another ordering $\sigma_k$ if possible, otherwise, continue.

\item Go to Step 4(b) and choose another basis if possible, otherwise, continue.

\item Go to Step 4(a) and choose another set of RDCSSs if possible, otherwise report that $d_1$ and $d_2$ are non-isomorphic.
\end{enumerate}
\end{algorithmic}
\end{algorithm}

\subsection{Isomorphism check for star based design}

A balanced star  $\Omega = St(n, \mu, t, t_{0})$ can be expressed as $\Omega = \psi \times \pi$, where $\psi$ is a $((t-t_0)-1)$-spread of $\mathcal{P}_{n-t_0}$, and $\pi$ is $(t_0 -1)$-dimensional subspace in  $\mathcal{P}_{n} \setminus \mathcal{P}_{n-t_0}$. Thus, the isomorphism check between two star-based designs $d_1$ and $d_2$ (with $(t_0 -1)$-dimensional nuclei) can be reduced to checking isomorphism between two $(h-1)$-spreads of $\mathcal{P}_{n-t_0}$ by iterating through the elements of $\mathcal{C}_{n-t_0}$ instead of $\mathcal{C}_{n}$. Algorithm 2 summarizes the steps of how Algorithm 1 can be used to search for an IEC between $\psi_1$ and $\psi_2$ assuming $\Omega_i = \psi_i \times \pi_i$ for $i=1,2$.\\

\begin{algorithm}[h!]
\caption{Isomorphism check between two $ St(n, \mu, t, t_{0})$-based designs $d_1$ and $d_2$, which correspond to stars $\Omega_1 = \psi_1 \times \pi_1$ and $\Omega_2 = \psi_2 \times \pi_2$, respectively.}
\begin{algorithmic}[1]

\item Determine two bases $\lbrace p_{1,1},\dots ,p_{1,t_0}\rbrace$ and $\lbrace p_{2,1},\dots ,p_{2,t_0}\rbrace$ of the nuclei $\pi_1$ and $\pi_2$, respectively.

\item Construct a collineation matrix $\mathcal{C}_{\pi_1 , B}$ mapping $\lbrace p_{1,1},\dots ,p_{1,t_0}\rbrace$ to the $t_0$ trailing basic factors $F_{n-t_0 +1},\dots, F_n$. The pre-images of $F_1, \dots , F_{n - t_0}$ can be chosen as an arbitrary linearly independent set from $\mathcal{P}_{n} \setminus \pi_1$.

\item Similarly, construct a collineation matrix $\mathcal{C}_{\pi_2 , B}$ mapping $\lbrace p_{2,1},\dots ,p_{2,t_0}\rbrace$  to the $t_0$ trailing basic factors $F_{n - t_0 +1},\dots, F_n$.

\item Extract designs $d_1^*$ and $d_2^*$ on $\mathcal{P}_{n-t_0}$ corresponding to the spreads $\mathcal{C}_{\pi_1 , B(\psi_1)}$ and $\mathcal{C}_{\pi_2 , B(\psi_2)}$.

\item Run Algorithm 1 on $d_1^*$ and $d_2^*$. If we come across a $\mathcal{C}^*$ that is an IEC, then an IEC for $d_1$ and $d_2$ is given by $\mathcal{C} = \mathcal{C}_{\pi_2 , B}^{-1} . \mathcal{C}^{*} . \mathcal{C}_{\pi_1 , B} $. Otherwise, $d_1$ and $d_2$ are non-isomorphic.

\end{algorithmic}
\end{algorithm}

\section{IsoCheck package} \label{sec:illustrations}

In this section, we examine the functions in the \pkg{IsoCheck} package which checks the isomorphism for RDCSS-based designs through the theoretical results and efficient algorithms proposed by \cite{spencer_etal_2019}. We will discuss both spread-based and star-based designs. The most important functions are \fct{checkSpreadIsomorphism} and  \fct{checkStarIsomorphism} which take two spreads and stars, respectively, as inputs and return a Boolean (TRUE/FALSE) indicating whether or not the inputs are isomorphic. If they are isomorphic, then these two functions also return isomorphism establishing collineations (IECs). We have also packaged a few important additional functions for efficiently checking the equivalence and iterating through different collineations. All of these functions have been summarized as follows:


\begin{itemize}
	\item \fct{checkSpreadIsomorphism}: requires two spreads  (\code{spread1} and \code{spread2}) as inputs whose isomorphism is to be checked.  It also takes an argument called \code{returnfirstIEC} which indicates whether all isomorphism establishing collineations should be returned (default), or terminate only after the first one is found.
	
	\item \fct{checkStarIsomorphism}: works in the similar way as \fct{checkSpreadIsomorphism}, except it requires two stars (\code{star1} and \code{star2}) as inputs, and uses Algorithm~2. 
	
  	\item \fct{checkSpreadEquivalence}: takes two spreads  (\code{spread1} and \code{spread2}) as inputs, and compares their sorted bitstring representations. It returns a Boolean- TRUE if the spreads are equivalent, else it returns FALSE.

	\item \fct{checkStarEquivalence} : works in the similar way as \fct{checkSpreadEquivalence}, except it requires two stars (\code{star1} and \code{star2}) as inputs.

  	\item \fct{getBitstrings} : takes a spread (or a star) and returns its bitstring representation. This makes the equivalence check, and subsequently the isomorphism check, significantly more computationally efficient as compared to the naive approach.
  
	\item \fct{applyCollineation} : takes an $n \times n$ collineation matrix and a spread (or a star), and produces the relabeled spread (or the star).
  
  \item \fct{star\_to\_spread} : takes a star \code{star} (assumes the structure to be $\Omega = \psi \times \pi$),  isolates the nucleus $\pi$ from this star by using a suitable collineation matrix, and returns the spread $\psi$ embedded in this star along with the collineation matrix. 
\end{itemize}

We have also included two more functions called \fct{stringtovector} and \fct{vectortostring} in the \proglang{R} package \pkg{IsoCheck}, which facilitates the users in inputting the spreads and stars. However, for quick usage and illustration, several spreads and stars have also been hard-coded, e.g., \code{spreadn4t2a, spreadn4t2b, starn8t5a, starn8t5b} (see the package manual for other builtin spreads and stars). These functions are illustrated in details in Section~4. \\

The function \fct{star\_to\_spread} is called within  \fct{checkStarIsomorphism} which separates the nucleus $\pi_{i}$ and 1($=(t-t_{0})-1$)-spread  $\psi_{i}$ of $\mathcal{P}_{4}$ from the star $\Omega_{i}$ where $i=1, 2$ such that $\Omega_{1} = \psi_{1} \times \pi_{1}$ and $\Omega_{2} = \psi_{2} \times \pi_{2}$ . Then the problem is reduced to checking the isomorphism between the two spreads  $\psi_{1}$ and  $\psi_{2}$ and equality of two nuclei $\pi_1$ and $\pi_2$. This is Algorithm 2 described in detail by \cite{spencer_etal_2019}. See Example~4 for an illustration.\\

\section{Examples using IsoCheck} \label{sec:examples}

This section demonstrates the usage of different functions in \pkg{IsoCheck} for checking equivalence and isomorphism of multi-stage factorial designs with randomization restrictions.  We also show how some of these functions can be used in design ranking and further creating catalogs of good non-isomorphic designs for a practitioner.\\

\textbf{Example 1} We know from \cite{andre1954nicht} that there exists non-trivial $1$-spreads $(t = 2)$ of $\mathcal{P}_{4}$ and \cite{soicher2000computation} has shown that they are isomorphic. Here we illustrate how isomorphism can be established between two $1$-spreads of $\mathcal{P}_{4}$. Table~3 shows the two $1$-spreads of $\mathcal{P}_{4}$. 

\begin{table}[!h]
	\caption{Two 1-spreads of $\mathcal{P}_{4} =  PG(3, 2)$}
	\label{Table 1}
	\begin{center}
		
		\begin{tabular}{@{}ccccclccccc@{}}
			\toprule
			\multicolumn{5}{c}{\code{spreadn4t2a}} &  & \multicolumn{5}{c}{\code{spreadn4t2b}} \\ \cmidrule(r){1-5} \cmidrule(l){7-11} 
			$f_{1}$ & $f_{2}$ & $f_{3}$ & $f_{4}$ & $f_{5}$ &  &$g_{1}$  & $g_{2}$ &  $g_{3}$ &  $g_{4}$ & $g_{5}$ \\ \cmidrule(r){1-5} \cmidrule(l){7-11} 
			D & C & B & A & CD &  & A & C & D & ABC & AC \\
			BC & AB & ACD & BD & AC &  & CD & ABCD & B & AD & AB \\
			BCD & ABC & ABCD & ABD & AD &  & ACD & ABD & BD & BCD & BC \\ \cmidrule(r){1-5} \cmidrule(l){7-11} 
			\bottomrule
		\end{tabular}
	\end{center}
\end{table}

The \proglang{R} package \pkg{IsoCheck} includes these two $1$-spreads of $\mathcal{P}_{4}$ that can be called using the \fct{data} command as follows:

\begin{CodeInput}
R> data("spreadn4t2a", package = "IsoCheck")
R> data("spreadn4t2b", package = "IsoCheck")
\end{CodeInput}

The users can use \fct{stringtovector} to manually input the spreads of their choice as a three dimensional array \code{spr[i, j, k]} which specifies the presence of $i$-th basic factor in the $j$-th effect of the $k$-th flat of \code{spr}. For instance, \code{spreadn4t2a} can be coded as follows.

\begin{CodeInput}
R> spr <- array(NA,c(n=4,flatsize=3,mu=5))
R> spr[,1,1] <- stringtovector("D",n=4)
R> spr[,2,1] <- stringtovector("BC",n=4)
R> spr[,3,1] <- stringtovector("BCD",n=4)
...
R> spr[,3,5] <- stringtovector("AD",n=4)	
\end{CodeInput}

We can check the isomorphism between the two spreads \code{spreadn4t2a} and \code{spreadn4t2b} (shown in Table~3) by calling  \fct{checkSpreadIsomorphism}. This function implements Algorithm~1 described in \cite{spencer_etal_2019} to search for the IEC by iterating through all the possible collineations over $\mathcal{P}_4$.

\begin{CodeInput}
R> test <- checkSpreadIsomorphism(spreadn4t2a, spreadn4t2b,
          +  returnfirstIEC = FALSE, printstatement = TRUE)
\end{CodeInput}

The output printed on console is shown as follows:

\begin{CodeOutput}
[1] "percent done: 0.28"
[1] "percent done: 0.56"
...
[1] "percent done: 99.72"
[1] "percent done: 100"
[1] "The two spreads are isomorphic. For example, one isomorphism establishing
 collineation is"
     [,1] [,2] [,3] [,4]
[1,]    0    1    0    0
[2,]    1    0    0    0
[3,]    1    1    1    1
[4,]    1    0    0    1
\end{CodeOutput}

The argument \code{printstatement} when set to \code{TRUE} declares the isomorphism at the end of the run and returns the first IEC. The output of the function \fct{checkSpreadIsomorphism} returns a Boolean value TRUE and enumerates all the isomorphism establishing collineations. The command \code{test\$IECs[[1]]} shows the first IEC, and   \code{length(test\$IECs)} returns the total number of IECs found by the algorithm. The user can instead choose to terminate the search for IEC as soon as the first IEC is found and cut down the run time, by using the argument \code{returnfirstIEC = TRUE}.\\

\code{R> test\$result}

\begin{CodeOutput}
[1] TRUE
\end{CodeOutput}

\code{R> test\$IECs[[1]]}

\begin{CodeOutput}
     [,1] [,2] [,3] [,4]
[1,]    0    1    0    0
[2,]    1    0    0    0
[3,]    1    1    1    1
[4,]    1    0    0    1
\end{CodeOutput}

\code{R> length(test\$IECs)}

\begin{CodeOutput}
[1] 360
\end{CodeOutput}

In order to cross-verify that the resultant matrix \code{test\$IEC[[1]]} indeed establishes isomorphism between the two spreads, we first apply the collineation matrix to the first spread \code{spreadn4t2a} and then check if the resulting spread \code{Cspreadn4t2a} is equivalent to the second spread \code{spreadn4t2b}. 

\begin{CodeInput}
	R> IEC <- test\$IECs[[1]]
	R> Cspreadn4t2a <- applyCollineation(IEC, spreadn4t2a)
	R> vectortostring(Cspreadn4t2a)
\end{CodeInput}

Table~4 presents the output of \code{vectortostring(Cspreadn4t2a)}. 

\begin{table}[!h]
	\caption{Equivalent of 1-spread of PG(3, 2)}
	\label{Table 2}
	\begin{center}
		\begin{tabular}{@{}ccccc@{}}
			\toprule
			\multicolumn{5}{c}{\code{Cspreadn4t2a}} \\ \midrule
			$h_{1}$ & $h_{2}$ & $h_{3}$ & $h_{4}$ & $h_{5}$ \\ \midrule
			CD & C & AC & BCD & D\\
			A & ABD & BC & AD & BD \\
			ACD & ABCD & ABC & ABC & B \\ \midrule
		\end{tabular}
	\end{center}
\end{table}

By visual inspection it is clear that \code{spreadn4t2b} is simply a rearrangement of \code{Cspreadn4t2a}, but for bigger problems it becomes difficult to conclude anything about equivalence by simple visual inspection. Hence we use \fct{checkSpreadEquivalence} function, which converts the spreads to their bitstring representations and the comparison becomes easy.

\begin{CodeInput}
R> checkSpreadEquivalence(Cspreadn4t2a, spreadn4t2b)
\end{CodeInput}
\begin{CodeOutput}
[1] TRUE
\end{CodeOutput}

\textbf{Example 2} We now consider the isomorphism of 2-spreads ($t = 3$) of $\mathcal{P}_{6} =  PG(5, 2)$. For illustration, we choose two distinct built-in spreads, \code{spreadn6t3a} and \code{spreadn6t3b}, however, the users may input the spreads of their own choice.

\begin{CodeInput}
R> checkSpreadEquivalence(spreadn6t3a, spreadn6t3b)
\end{CodeInput}
\begin{CodeOutput}
[1] FALSE
\end{CodeOutput}

%
%
The two spreads \code{spreadn6t3a} and \code{spreadn6t3b} are presented in Tables~5 and 6 respectively. \\

\begin{table}[!h]
\caption{A 2-spread of $\mathcal{P}_{6} =  PG(5, 2)$}
\label{Table 3}
\begin{center}
\begin{tabular}{@{}ccccccccc@{}}
\toprule
\multicolumn{9}{c}{\code{spreadn6t3a}} \\ \midrule
$f_{1}$ & $f_{2}$ & $f_{3}$ & $f_{4}$ & $f_{5}$ &  $f_{6}$ & $f_{7}$ & $f_{8}$ & $f_{9}$\\ \midrule
A	&	B	&	C	&	D	&	E	&	F	&	BD	&	AC	&	AD	\\
EF	&	AF	&	AB	&	BC	&	CD	&	DE	&	BF	&	CE	&	BE	\\
AEF	&	ABF	&	ABC	&	BCD	&	CDE	&	DEF	&	DF	&	AE	&	ABDE	\\
BCE	&	CDF	&	ADE	&	BEF	&	ACF	&	ABD	&	ACE	&	BDF	&	CF	\\
ABCE	&	BCDF	&	ACDE	&	BDEF	&	ACEF	&	ABDF	&	ABCDE	&	ABCDF	&	ACDF	\\
BCF	&	ACD	&	BDE	&	CEF	&	ADF	&	ABE	&	ABCEF	&	BCDEF	&	BCEF	\\
ABCF	&	ABCD	&	BCDE	&	CDEF	&	ADEF	&	ABEF	&	ACDEF	&	ABDEF	&	ABCDEF	\\ \midrule

\end{tabular}
\end{center}
\end{table}

\begin{table}[!h]
\caption{A 2-spread of $\mathcal{P}_{6} =  PG(5, 2)$}
\label{Table 4}
\begin{center}
\begin{tabular}{@{}ccccccccc@{}}
\toprule
\multicolumn{9}{c}{\code{spreadn6t3b}} \\ \midrule
$g_{1}$ & $g_{2}$ & $g_{3}$ & $g_{4}$ & $g_{5}$ &  $g_{6}$ & $g_{7}$ & $g_{8}$ & $g_{9}$\\ \midrule
ABC	&	E	&	DF	&	B	&	A	&	EF	&	BE	&	ABCDF	&	AC	\\
AEF	&	ABCEF	&	ABCE	&	DEF	&	BDF	&	AB	&	F	&	ADF	&	AE	\\
BCEF	&	ABCF	&	ABCDEF	&	BDEF	&	ABDF	&	ABEF	&	BEF	&	BC	&	CE	\\
ADEF	&	BDE	&	C	&	AF	&	ABCDE	&	ACE	&	BCDF	&	BF	&	DE	\\
BCDEF	&	BD	&	CDF	&	ABF	&	BCDE	&	ACF	&	CDEF	&	ACD	&	ACDE	\\
D	&	ACDF	&	ABE	&	ADE	&	ACEF	&	BCE	&	BCD	&	ABD	&	AD	\\
ABCD	&	ACDEF	&	ABDEF	&	ABDE	&	CEF	&	BCF	&	CDE	&	CF	&	CD	\\ \midrule

\end{tabular}
\end{center}
\end{table}

We now proceed to check isomorphism between the two spreads.

\begin{CodeInput}
	R> test <- checkSpreadIsomorphism(spreadn6t3a, spreadn6t3b,
	+   returnfirstIEC = TRUE, printstatement = FALSE)
\end{CodeInput}

\code{R> test\$result}

\begin{CodeOutput}
	[1] TRUE
\end{CodeOutput}

\code{R> test\$IECs}

\begin{CodeOutput}
	[[1]]
	[,1] [,2] [,3] [,4] [,5] [,6]
	[1,]    1    0    0    1    1    0
	[2,]    0    1    0    1    0    0
	[3,]    1    1    0    0    0    1
	[4,]    1    1    1    0    0    1
	[5,]    0    1    0    1    0    1
	[6,]    1    1    1    1    1    0
\end{CodeOutput}

Similar to Example~1, one can use \fct{applyCollineation} and \fct{checkSpreadEquivalence} to cross-check whether the collineation matrix found is indeed an IEC. Since we used "\code{returnfirstIEC = TRUE}", the code terminates after the first IEC was found. \\

\textbf{Example 3} 
\cite{mee_bates_1998} constructed and analyzed split-lot experimental designs for multistage batch processing for fabrication of integrated circuits. Here we illustrate the isomorphism between two designs for a 64-wafer experiment which involves six basic factors and nine disjoint RDCCSs, each of size seven. The two designs are $IC_{1}= \lbrace  \langle A, EF, BCE\rangle$, $\langle B, AF, CDF\rangle$, $\langle C, AB, ADE\rangle$, $\langle D, BC, BEF\rangle$, $\langle E, CD, ACF\rangle$, $\langle F, DE, ABD\rangle$, $\langle BD, BF, ACE\rangle$, $\langle AC, CE, BDF\rangle$, $\langle AD, BE, CF\rangle \rbrace$ and $IC_{2}= \lbrace  \langle A, BD, CF\rangle$, $\langle B, AF, CE\rangle$, $\langle C, BF, DE\rangle$, $\langle D, AC, BE\rangle$, $\langle E, AB, DF\rangle$, $\langle F, AE, CD\rangle$, $\langle AD, BC, EF\rangle$, $\langle ACE, ADF, BEF\rangle$, $\langle ABC, ADE, CEF\rangle \rbrace$. Here, $\langle .,.,.\rangle$ denotes the span of the basis vectors within it. We can construct the full flats using these basis vectors with the following code.

\begin{CodeInput}
R> basistoflat3 <- function(b1,b2,b3, n){
+  base1 <- stringtovector(b1, n)
+  base2 <- stringtovector(b2, n)
+  base3 <- stringtovector(b3, n)
+  v1  <- (base1 + base2) %%2
+  v2  <- (base1 + base3) %%2
+  v3  <- (base2 + base3) %%2
+  v4  <- (v1 + base3) %%2
+  return(cbind(base1, base2, v1, base3, v2, v3,v4))
+ }

R> n <- 6
R> flatsize <- 2^3-1
R> mu <- 9
R> spread_IC1 <- array(NA, c(n, flatsize, mu))

R> spread_IC1[,,1] <- basistoflat3("A", "EF", "BCE", n)
R> spread_IC1[,,2] <- basistoflat3("B", "AF", "CDF", n)
...
R> spread_IC1[,,9] <- basistoflat3("AD", "BE", "CF", n)
\end{CodeInput}

The complete spread $IC_{1}$ is now constructed and stored in the array \code{spread_IC1}. Similarly, the second spread $IC_{2}$ is constructed and stored as \code{spread_IC2}.  We can now proceed to check the isomorphism between \code{spread_IC1} and \code{spread_IC2}.

\begin{CodeInput}
R> test<- checkSpreadIsomorphism(spread_IC1, spread_IC2, 
+ returnfirstIEC = T)
\end{CodeInput}
\code{R> test\$result}

\begin{CodeOutput}
[1] TRUE
\end{CodeOutput}

Hence, both the designs $IC_{1}$ and $IC_{2}$ are isomorphic. As in Example~2, we saved some computational time by terminating the algorithm after the first IEC was found. One can extract the IEC by calling \code{test\$IECs}.\\

\textbf{Example 4}  We now demonstrate the isomorphism check between two stars \code{starn5t3a}  and \code{starn5t3b} denoted by $\Omega_{1}$ and $\Omega_{2}$, respectively, (see Table~7). Both of these are balanced covering stars of $\mathcal{P}_{5} =  PG(4, 2)$ denoted by $ St(5, 5, 3, 1)$. The stars consist of $\mu= \frac{2^{n-t_{0}}-1}{2^{t-t_{0}}-1}= \frac{2^{5-1}-1}{2^{3-1}-1} = 5$ rays of 2-flats overlapping with a nucleus which is a 0-flat in $\mathcal{P}_{5}$. A close look at Table~7 shows that the nucleus in $\Omega_{1}$ is $\pi_{1}= \langle A \rangle$ and in $\Omega_{2}$ it is $\pi_{2}= \langle ABC \rangle$. \\

\begin{table}[!h]
\caption{Two stars of $\mathcal{P}_{5} =  PG(4, 2)$}
\label{Table 6}
\begin{center}
\begin{tabular}{@{}ccccclccccc@{}}
\toprule
\multicolumn{5}{c}{\code{starn5t3a}$= \Omega_{1}$} &  & \multicolumn{5}{c}{\code{starn5t3b}$= \Omega_{2}$} \\ \cmidrule(r){1-5} \cmidrule(l){7-11} 
$f_{1}$ & $f_{2}$ & $f_{3}$ & $f_{4}$ & $f_{5}$ &  &$g_{1}$  & $g_{2}$ &  $g_{3}$ &  $g_{4}$ & $g_{5}$ \\ \cmidrule(r){1-5} \cmidrule(l){7-11} 
A	&	D	&	C	&	B	&	DE	&	&	ABC	&	AE	&	D	&	E	&	CE	\\
E	&	BC	&	BDE	&	BCE	&	BD	&	&	AC	&	DE	&	C	&	ACDE	&	A	\\
CDE	&	BCD	&	BCDE	&	CE	&	BE	&	&	CDE	&	AD	&	CD	&	ACD	&	ACE	\\
AE	&	AD	&	AC	&	AB	&	A	&	&	B	&	BCE	&	ABCD	&	ABCE	&	ABC	\\
ACD	&	ABC	&	ABDE	&	ACE	&	ABD	&	&	BCDE	&	ABCDE	&	AB	&	BD	&	BC	\\
ACDE	&	ABCD	&	ABCDE	&	A	&	ABE	&	&	ABDE	&	BCD	&	ABD	&	ABC	&	BE	\\
CD	&	A	&	A	&	ABCE	&	ADE	&	&	ADE	&	ABC	&	ABC	&	BDE	&	ABE	\\
\cmidrule(r){1-5} \cmidrule(l){7-11} 
  \bottomrule
\end{tabular}
\end{center}
\end{table}

The following code can be used to quickly verify the equivalence of the two stars.

\begin{CodeInput}
R> checkStarEquivalence(starn5t3a, starn5t3b)
\end{CodeInput}
\begin{CodeOutput}
[1] FALSE
\end{CodeOutput}

As discussed in Sections~2, one can check for the isomorphism between two stars either by directly using \fct{checkStarIsomorphism} or by first decomposing the star to a spread via \fct{star\_to\_spread} and then use \fct{checkSpreadIsomorphism}. The first approach is implemented as follows.

\begin{CodeInput}
R> checkStarIsomorphism(starn5t3a, starn5t3b, returnfirstIEC = TRUE)
\end{CodeInput}
\begin{CodeOutput}
$result
[1] TRUE

$IECs[[1]]
X1 X2 X3 X4 X5
X1  1  0  1  1  1
X2  1  1  1  0  0
X3  1  0  0  0  0
X4  0  1  0  0  1
X5  0  1  1  1  1
\end{CodeOutput}

For the latter approach, we first assume that $\Omega_i = \psi_i \times \pi_i$, for $i=1,2$. Then one can extract $\psi_1$ and $\psi_2$ using the function \fct{star\_to\_spread} on \code{starn5t3a} and \code{starn5t3b}, respectively. 

\begin{CodeInput}
R> red_spread_1 <- star_to_spread(starn5t3a)
R> red_spread_2 <- star_to_spread(starn5t3b)
R> psi_1 <- vector_to_string(red_spread_1[[1]])
R> psi_1
\end{CodeInput}

\begin{CodeOutput}
	[,1]  [,2]  [,3]   [,4]  [,5]
	[1,] "A"   "B"   "C"    "D"   "AB"
	[2,] "ABC" "CD"  "ABD"  "ACD" "BD"
	[3,] "BC"  "BCD" "ABCD" "AC"  "AD"
\end{CodeOutput}

\begin{CodeInput}
R> psi_2 <- vector_to_string(red_spread_2[[1]])
R> psi_2
\end{CodeInput}

\begin{CodeOutput}
	[,1]   [,2] [,3]  [,4]  [,5] 
	[1,] "C"    "AD" "B"   "A"   "D"  
	[2,] "ABCD" "AB" "CD"  "BC"  "AC" 
	[3,] "ABD"  "BD" "BCD" "ABC" "ACD"
\end{CodeOutput}

The output of \fct{star\_to\_spread} is a list containing two values, the reduced spread and a collineation matrix over $\mathcal{P}_5$ that ensures that the spreads $\psi_i$ spans $\langle A,B,C,D\rangle$ and the nucleus is $\pi_i = \langle E \rangle$. For instance, the collineation matrix for relabelling $\Omega_2$ is obtained as

\begin{CodeInput}
R> red_spread_2$collineation
\end{CodeInput}

\begin{CodeOutput}
   X1 X2 X3 X4 X5
    0  0  0  0  1
    0  0  0  1  0
    0  1  1  0  0
    1  0  1  0  0
    0  0  1  0  0
\end{CodeOutput}

and the corresponding relabelled star can be extracted as follows.

\begin{CodeInput}
R> Cstarn5t3b <- applyCollineation(red_spread_2$collineation,starn5t3b)
R> vectortostring(Cstarn5t3b)
\end{CodeInput}

\begin{CodeOutput}	
     [,1]    [,2]  [,3]   [,4]   [,5]  
[1,] "E"     "AD"  "B"    "A"    "ACDE"
[2,] "CE"    "AB"  "CDE"  "ABCE" "D"   
[3,] "ABCDE" "BD"  "BCDE" "BCE"  "ACE" 
[4,] "C"     "ADE" "BE"   "AE"   "E"   
[5,] "ABDE"  "ABE" "CD"   "BC"   "DE"  
[6,] "ABCD"  "BDE" "BCD"  "E"    "AC"  
[7,] "ABD"   "E"   "E"    "ABC"  "ACD" 
\end{CodeOutput}

We now check the isomorphism between $\Omega_1$ and $\Omega_2$ by simply checking the isomorphism between \code{red_spread_1[[1]]} and \code{red_spread_2[[1]]}.

\begin{CodeInput}
R> checkSpreadIsomorphism(red_spread1[[1]], red_spread2[[1]], 
	+ returnfirstIEC=TRUE)
\end{CodeInput}

\begin{CodeOutput}
$result
[1] TRUE

$IECs[[1]]
     [,1] [,2] [,3] [,4]
[1,]    1    1    1    1
[2,]    1    0    0    1
[3,]    0    0    1    1
[4,]    1    1    1    0
\end{CodeOutput}

\textbf{Example 5} \cite{bingham2008factorial} considered a $2^{5}$ factorial split-lot design for manufacturing plutonium alloy cookies which involved five basic factors with three processing stages. They suggested a design $PA_{1}= \lbrace \langle A, B, CDE \rangle, \langle C, AD, BE\rangle, \langle D, E, ABC \rangle\rbrace$. This design is derived from a balanced covering star  $ St(5, 5, 3, 1)$ of $\mathcal{P}_{5}$. Later on, Ranjan et al. (2010) recommended another design $PA_{2}= \lbrace\langle A, B, DE, ACD \rangle, \langle C, AB, DE, ACD \rangle, \langle D, E, AB, ACD \rangle\rbrace$, which represents a balanced covering star  $ St(5, 3, 4, 3)$ of $\mathcal{P}_{5}$. 

We can define the two stars as shown in the following code. For defining $PA_1$,  \fct{basistoflat3} was presented in Example~3. We similarly write a function called \fct{basistoflat4} for constructing the $(4-1)$-flats, and then define $PA_2$ as follows.

\begin{CodeInput}
R> n <- 5
R> flatsize <- 2^3-1
R> mu <- 3	

R> star_PA1 <- array(NA, c(n, flatsize, mu))
R> star_PA1[,,1] <- basistoflat3("A", "B", "CDE", n)
R> star_PA1[,,2] <- basistoflat3("C", "AD", "BE", n)
R> star_PA1[,,3] <- basistoflat3("D", "E", "ABC", n)

R> n <- 5
R> flatsize <- 2^4-1
R> mu <- 3	

R> star_PA2 <- array(NA, c(n, flatsize, mu))
R> star_PA2[,,1] <- basistoflat4("A", "B", "DE", "ACD", n)
R> star_PA2[,,2] <- basistoflat4("C", "AB", "DE", "ACD", n)
R> star_PA2[,,3] <- basistoflat4("D", "E", "AB", "ACD", n)
\end{CodeInput}

The isomorphism between the two stars can be checked by using the following code.
\begin{CodeInput}
R> checkStarIsomorphism(star_PA1, star_PA2)
\end{CodeInput}
\begin{CodeOutput}
Stars are not of same dimension.
[1] FALSE
\end{CodeOutput}

The output is certainly expected as the \code{flatsize} in $PA_1$ and $PA_2$ are different. One can also verify that the sizes of nucleus in these two stars are different.

\begin{CodeInput}
R> nucleus_PA1 <- suppressMessages(plyr::match_df(data.frame(t(star_PA1[,,1])),
 + data.frame(t(star_PA1[,,2]))))
R> vectortostring(as.numeric(nucleus_PA1))
\end{CodeInput}
\begin{CodeOutput}
     [,1]   
[1,] "ABCDE"
\end{CodeOutput}

\begin{CodeInput}
R> nucleus_PA2 <- suppressMessages(plyr::match_df(data.frame(t(star_PA2[,,1])),
 +  data.frame(t(star_PA2[,,2]))))
R> vectortostring(t(as.matrix(nucleus_PA2)))
\end{CodeInput}

\begin{CodeOutput}
     [,1]  
[1,] "DE"  
[2,] "ACD" 
[3,] "AB"  
[4,] "ACE" 
[5,] "ABDE"
[6,] "BCD" 
[7,] "BCE" 
\end{CodeOutput}

\section{Catalog of Multi-Stage Factorial Designs}\label{sec:catalog}

In this section, we illustrate the usage of \pkg{IsoCheck} functions in catalog construction for balanced spread- and balanced covering star- based designs. We do not claim to propose any new designs, new ranking criterion, or new catalogs. In general, the construction of such a catalog requires iterating through all possible candidate designs and then checking isomorphism among them \citep{bingham1999minimum, lin_sitter_2008}. Unfortunately, we are unaware of any computationally efficient methodologies that can construct all possible balanced $(t-1)$- spreads or balanced covering stars of $PG(n-1, 2)$. As a result, we follow an exhaustive search approach for constructing all possible spreads.\\

Researchers and practitioners often combine such catalogs with a carefully chosen design ranking criteria to sort through different designs and select the most optimal candidate for their experiments.  For classical fractional factorial designs, popular design ranking criteria  include maximum resolution, generalized minimum aberration, maximum number of clear two-factor interactions (2-fi's), etc. \citep{xu2001generalized, butler2003some, cheng2011paper}. Let $d$ be a  factorial design with randomization restrictions characterized by $S_1,...,S_k$,  then, \cite{bingham2008factorial} proposed a V-criterion defined by 
$$ V(d) = \sum_{i=1}^k (p_i-\bar{p})^2, $$
where $\bar{p}=\sum_{i=1}^k p_i / k$ and $p_i$ is proportion of main effects and 2-fi's in $S_i$. If we have a $(t-1)$-spread based design then $S_i$ would an RDCSS, or equivalently, a $(t-1)$-flat in the $(t-1)$-spread, whereas, if the design is obtained from a star $St(n, \mu, t, t_{0})$, then $S_i$ would be the flat minus the nucleus (i.e., $|S_i|=2^t-2^{t_0}$). From an analysis standpoint,  if we use half-normal plots for assessing the significance of factorial effects, then  $S_i$ represents the set of factorial effects that would be used for the $i$-th half-normal plot. Thus minimizing this $V$-criterion encourages homogeneous distribution of main effects and two-factor interactions among the $S_i$'s.  
\\

We now discuss building a catalog of designs characterized by a $(t-1)$-spread of $\mathcal{P}_n$, for fixed $t<n$.  Suppose we wish to build a catalog of $1$-spreads of $\mathcal{P}_4=PG(3,2)$ (Table~3 presents an example \code{spreadn4t2a}).  This structure contains $2^4-1=15$ elements (factorial effects) distributed in 5 subspaces of size 3 each.  One possible (although computationally inefficient) approach is to exhaustively search all possible permutations/arrangements of these 15 elements, and then check whether or not they form a $1$-spread.  If yes, then we can check the isomorphism between this spread and the original \code{spreadn4t2a}. We can also compute the design ranking criterion value for all these spreads. The steps of catalog construction are summarized in Algorithm~3.\\

\begin{algorithm}[h!]
	\caption{for constructing a catalog of non-isomorphic spreads (or stars).}
	\begin{algorithmic}[1]
		
		\item Pick an initial $(t-1)$-spread (say, $d_1$) of $\mathcal{P}_n := PG(n-1,2)$.
		
		\item Compute the design ranking criterion value of $d_1$.
		
		\item Pick a random permutation (say $\rho$) of the set of $2^n-1$ elements of $d_1$.
		\item Use \code{is.spread()} to check if $\rho(d_1)$ is a proper $(t-1)$-spread of $\mathcal{P}_n$. If ``TRUE", go to the next step, otherwise go back to Step 3 (if all possible permutations of $d_1$ have not been exhausted yet).
		
		\item Compute the design ranking criterion value of $\rho(d_1)$.
		\item Use \code{checkSpreadIsomorphism()} to check whether or not $\rho(d_1)$ is isomorphic to $d_1$.
		
		\item Go to Step 3 if all possible permutations of $d_1$ have not been exhausted yet.
		
	\end{algorithmic}
\end{algorithm}

Note that $15! = 1.307674 \times 10^{12}$  is a big number and one would have to iterate through these many possible rearrangements to ensure the construction of all possible $1$-spreads of $\mathcal{P}_4$. Admittedly, one can come up with innovative ideas to save some computational cost, but that is not the main objective here. For this example, we ran this algorithm for over 5 million random permutations, and found almost 110 isomorphic spreads. Note that we did not expect to find any non-isomorphic spreads, because  \cite{spencer_etal_2019} have already mentioned that all $1$-spreads of $PG(3,2)$ are isomorphic. Nonetheless, we used this spread for illustration because it is a computationally inexpensive small example to follow through. Interestingly, we found designs with two criterion value $V(d)=0.44$ and $0.22$, which indicates preference of one design over the other. The initial design  \code{spreadn4t2a} gave $V(d)=0.22$ (with $p_1 = 2/3, p_2=2/3, p_3=1/3, p_4=2/3$ and $p_5=1$), whereas, the following design, found through Algorithm~3, corresponds to $V(d)=0.44$ (with $p_1 = 1/3, p_2=2/3, p_3=1, p_4=1/3$ and $p_5=1$).	

\begin{CodeOutput}	
	[,1]  [,2]  [,3]  [,4]  [,5]  
	[1,] "ABC"  "CD"  "AD" "ABD" "B" 
	[2,] "D"    "ACD" "AB" "BCD" "C" 
	[3,] "ABCD" "A"   "BD" "AC"  "BC"
\end{CodeOutput}


If the underlying design is based on a star (say $\Omega$), then one can use \code{star\_to\_spread}($\Omega$) function to extract the spread $\psi$ after removing the nucleus $\pi$ (assuming $\Omega = \psi \times \pi$) from the star. Subsequently, the designs from the catalog of the underlying spreads can be combined with the nucleus to construct the desired catalog of stars. For instance, \code{star\_to\_spread(starn5t3a)}  gives a $1$-spread of $PG(3,2)$, which has been discussed in details.  \\


\section{Concluding Remarks} \label{sec:summary}
This paper presents an \proglang{R} package \pkg{IsoCheck} to check the isomorphism of balanced $(t-1)$-spreads and balanced covering stars of $\mathcal{P}_n=PG(n-1,2)$, which corresponds to the isomorphism check of two-level multi-stage factorial designs with randomization restrictions  \citep{ranjan2009existence, ranjan2010stars}. The main ideas behind the algorithms are to apply the relabellings via collineation matrices and rearrangements or the factorial effects within the RDCSSs through  bitstring representation of the elements of $\mathcal{P}_n$.  In this paper, we have presented several worked out examples which illustrate the usage of \pkg{IsoCheck} functions and their arguments for the isomorphism check. More importantly, we also outline how \pkg{IsoCheck} functions can be used to build catalog of good designs in this class of spread/star-based factorial designs.\\

A few important remarks are in order.
1. In practice, whenever we have more than one possible design options for conducting an experiment, the ranking of design becomes important. This has traditionally been done in the context of ``fractional" factorial designs, as we have several options for choosing the fraction defining contrast subgroups (FDCS). For multi-stage factorial designs with randomization restrictions (i.e., spread/star -based designs), we have to choose even randomization defining contrast subspaces (RDCSSs), which are in some sense similar to FDCS. Therefore, we can also talk about the isomorphism and ranking of the full factorial unreplicated spread/star-based multi-stage factorial designs with randomization restrictions. 2. One can use \pkg{IsoCheck} functions to check for the isomorphism of multi-stage fractional factorial designs as well. 3. The generalization of these isomorphism check algorithms for $q (\ge 3)$-level factorial designs appears to be non-trivial.




\section*{Acknowledgments}


\bibliography{refs_IsoCheck}

\begin{thebibliography}{29}
\newcommand{\enquote}[1]{``#1''}
\providecommand{\natexlab}[1]{#1}
\providecommand{\url}[1]{\texttt{#1}}
\providecommand{\urlprefix}{URL }
\expandafter\ifx\csname urlstyle\endcsname\relax
  \providecommand{\doi}[1]{doi:\discretionary{}{}{}#1}\else
  \providecommand{\doi}{doi:\discretionary{}{}{}\begingroup
  \urlstyle{rm}\Url}\fi
\providecommand{\eprint}[2][]{\url{#2}}

\bibitem[{Addelman(1964)}]{addelman1964}
Addelman S (1964).
\newblock \enquote{Some two-level fractional plans with split-plot
  confounding.}
\newblock \emph{Technometrics}, \textbf{30}, 253--258.

\bibitem[{Andr{\'e}(1954)}]{andre1954nicht}
Andr{\'e} J (1954).
\newblock \enquote{{\"U}ber nicht-desarguessche Ebenen mit transitiver
  Translationsgruppe.}
\newblock \emph{Mathematische Zeitschrift}, \textbf{60}(1), 156--186.

\bibitem[{Bailey(2004)}]{bailey2004association}
Bailey RA (2004).
\newblock \emph{Association schemes: Designed experiments, algebra and
  combinatorics}, volume~84.
\newblock Cambridge University Press.

\bibitem[{Batten(1997)}]{batten1997combinatorics}
Batten LM (1997).
\newblock \emph{Combinatorics of finite geometries}.
\newblock Cambridge University Press.

\bibitem[{Bingham and Sitter(1999{\natexlab{a}})}]{bingham_sitter_1999}
Bingham D, Sitter R (1999{\natexlab{a}}).
\newblock \enquote{Minimum-aberration two-level fractional factorial split-plot
  designs.}
\newblock \emph{Technometrics}, \textbf{41}, 62--70.

\bibitem[{Bingham \emph{et~al.}(2008)Bingham, Sitter, Kelly, Moore, and
  Olivas}]{bingham2008factorial}
Bingham D, Sitter R, Kelly E, Moore L, Olivas JD (2008).
\newblock \enquote{Factorial designs with multiple levels of randomization.}
\newblock \emph{Statistica Sinica}, pp. 493--513.

\bibitem[{Bingham and Sitter(1999{\natexlab{b}})}]{bingham1999minimum}
Bingham D, Sitter RR (1999{\natexlab{b}}).
\newblock \enquote{Minimum-aberration two-level fractional factorial split-plot
  designs.}
\newblock \emph{Technometrics}, \textbf{41}(1), 62--70.

\bibitem[{Bisgaard(1994)}]{bisgaard_1994}
Bisgaard S (1994).
\newblock \enquote{Blocking generators for small $2^{k-p}$ designs.}
\newblock \emph{Journal of Quality Technology}, \textbf{26}, 288--296.

\bibitem[{Bose(1947)}]{bose1947mathematical}
Bose RC (1947).
\newblock \enquote{Mathematical theory of the symmetrical factorial design.}
\newblock \emph{Sankhy{\=a}: The Indian Journal of Statistics}, pp. 107--166.

\bibitem[{Box \emph{et~al.}(1978)Box, Hunter, Hunter
  \emph{et~al.}}]{box1978statistics}
Box GE, Hunter WG, Hunter JS, \emph{et~al.} (1978).
\newblock \enquote{Statistics for experimenters.}

\bibitem[{Butler(2003)}]{butler2003some}
Butler NA (2003).
\newblock \enquote{Some theory for constructing minimum aberration fractional
  factorial designs.}
\newblock \emph{Biometrika}, \textbf{90}(1), 233--238.

\bibitem[{Cheng and Tsai(2011)}]{cheng2011paper}
Cheng CS, Tsai PW (2011).
\newblock \enquote{Multistratum Fractional Factorial Designs.}
\newblock \emph{Statistica Sinica}, \textbf{21}, 1001--1021.

\bibitem[{Fisher(1942)}]{Fisher1942}
Fisher RA (1942).
\newblock \enquote{The theory of confounding in factorial experiments in
  relation to the theory of groups.}
\newblock \emph{Annals of Eugenics}, \textbf{11}, 290--299.

\bibitem[{Franklin and Bailey(1977)}]{franklin_bailey_1977}
Franklin M, Bailey R (1977).
\newblock \enquote{Selection of defining contrasts and confounded effects in
  two-level experiments.}
\newblock \emph{Applied Statistics}, \textbf{26}, 321--326.

\bibitem[{Hirschfeld(1998)}]{hirschfeld1998projective}
Hirschfeld J (1998).
\newblock \emph{Projective Geometries Over Finite Fields. Oxford Mathematical
  Monographs}.
\newblock Oxford University Press New York.

\bibitem[{Lin and Sitter(2008)}]{lin_sitter_2008}
Lin C, Sitter R (2008).
\newblock \enquote{n Isomorphism Check for Two-Level Fractional Factorial
  Designs.}
\newblock \emph{Journal of Statistical Planning and Inference}, \textbf{134},
  1085--1101.

\bibitem[{Mee and Bates(1998)}]{mee_bates_1998}
Mee R, Bates R (1998).
\newblock \enquote{Split-lot designs: Experiments for multistage batch
  processes.}
\newblock \emph{Technometrics}, \textbf{40}, 127--140.

\bibitem[{Miller(1997)}]{miller_1997}
Miller A (1997).
\newblock \enquote{Strip-plot configurations of fractional factorials.}
\newblock \emph{Technometrics}, \textbf{39}, 153--161.

\bibitem[{Nelder(1965{\natexlab{a}})}]{nelder_1965a}
Nelder J (1965{\natexlab{a}}).
\newblock \enquote{The analysis of randomized experiments with orthogonal block
  structure. I. block structure and the null analysis of variance.}
\newblock \emph{Proc. R. Soc. Lond. A}, \textbf{283}, 147--162.

\bibitem[{Nelder(1965{\natexlab{b}})}]{nelder_1965b}
Nelder J (1965{\natexlab{b}}).
\newblock \enquote{The analysis of randomized experiments with orthogonal block
  structure. II. Treatment structure and the general analysis of variance.}
\newblock \emph{Proc. R. Soc. Lond. A}, \textbf{283}, 163--178.

\bibitem[{{R Core Team}(2020)}]{R-cran}
{R Core Team} (2020).
\newblock \emph{R: A Language and Environment for Statistical Computing}.
\newblock R Foundation for Statistical Computing, Vienna, Austria.
\newblock \urlprefix\url{https://www.R-project.org/}.

\bibitem[{Ranjan \emph{et~al.}(2010)Ranjan, Bingham, and
  Mukerjee}]{ranjan2010stars}
Ranjan P, Bingham D, Mukerjee R (2010).
\newblock \enquote{Stars and regular fractional factorial designs with
  randomization restrictions.}
\newblock \emph{Statistica Sinica}, pp. 1637--1653.

\bibitem[{Ranjan \emph{et~al.}(2009)Ranjan, Bingham, and
  Dean}]{ranjan2009existence}
Ranjan P, Bingham DR, Dean AM (2009).
\newblock \enquote{Existence and construction of randomization defining
  contrast subspaces for regular factorial designs.}
\newblock \emph{The Annals of Statistics}, pp. 3580--3599.

\bibitem[{Soicher(2000)}]{soicher2000computation}
Soicher L (2000).
\newblock \enquote{Computation of Partial Spreads, web preprint.}

\bibitem[{Speed and Bailey(1982)}]{speed_bailey_1982}
Speed TP, Bailey RA (1982).
\newblock \enquote{On a class of association schemes derived from lattices of
  equivalence relations.}
\newblock \emph{Algebraic Structures and Applications, Marcel Dekker, New
  York}, pp. 55--74.

\bibitem[{Spencer \emph{et~al.}(2019)Spencer, Ranjan, and
  Mendivil}]{spencer_etal_2019}
Spencer N, Ranjan P, Mendivil F (2019).
\newblock \enquote{Isomorphism Check for $2^n$ Factorial Designs with
  Randomization Restriction.}
\newblock \emph{Journal of Statistical Theory and Practice}, \textbf{13}(60),
  1--24.

\bibitem[{Tjur(1984)}]{tjur_1984}
Tjur T (1984).
\newblock \enquote{Analysis of variance models in orthogonal designs.}
\newblock \emph{Int. Statist. Rev.}, \textbf{52}, 33--81.

\bibitem[{Xu \emph{et~al.}(2001)Xu, Wu \emph{et~al.}}]{xu2001generalized}
Xu H, Wu CJ, \emph{et~al.} (2001).
\newblock \enquote{Generalized minimum aberration for asymmetrical fractional
  factorial designs.}
\newblock \emph{The Annals of Statistics}, \textbf{29}(4), 1066--1077.

\bibitem[{Yates(1937)}]{yates1937}
Yates F (1937).
\newblock \emph{Design and Analysis of Factorial Experiments}.
\newblock Technical Communication 35, Harpenden, United Kingdom: Imperial
  Bureau of Soil Science.

\end{thebibliography}







\end{document}